\documentclass[twocolumn, prl]{revtex4}
\usepackage{graphicx}
\usepackage{latexsym}

\newcommand{\be}{\begin{equation}}
\newcommand{\ee}{\end{equation}}
\newcommand{\bea}{\begin{eqnarray}}
\newcommand{\eea}{\end{eqnarray}}

\begin{document}

\title{How does a protein search for the specific
site on DNA: the role of disorder}
\author{Tao Hu, B. I. Shklovskii}
\affiliation{Department of Physics, University of Minnesota \\
116 Church Street SE, Minneapolis, MN 55455}
\date{\today}

\begin{abstract}

Proteins can locate their specific targets on DNA up to two orders
of magnitude faster than the Smoluchowski three-dimensional
diffusion rate. This happens due to non-specific adsorption of
proteins to DNA and subsequent one-dimensional sliding along DNA. We
call such one-dimensional route towards the target "antenna". We
studied the role of the dispersion of nonspecific binding energies
within the antenna due to quasi random sequence of natural DNA.
Random energy profile for sliding proteins slows the searching rate
for the target. We show that this slowdown is different for the
macroscopic and mesoscopic antennas.

\end{abstract}
\maketitle

A protein binding to a specific site on DNA, which we call the
target, is one of the central paradigms of biology~\cite{Gann}. Well
known examples include \emph{lac}-repressor in \emph{E. coli}, which
regulates a specific gene producing enzyme consuming lactose and the
proper restriction enzyme destroying genome of invading \emph{E.
coli} $\lambda$-phage in real time warfare for bacteria survival. It
is known since the early days of molecular biology that in some
cases proteins can find their target sites along a DNA chain one to
two orders faster than the maximum rate achievable by
three-dimensional diffusion~\cite{Riggs,Richter}. To resolve this
paradox, nonspecific binding and subsequent one-dimensional sliding
of proteins along the DNA to the target was suggested as an
important component of the searching process~\cite{Riggs, Richter}.
This idea was studied in various models proposed by both physicists
and biologists~\cite{BWH,Bruinsma,Marko,Halford,Klenin}. A
comprehensive study of interplay between the 1D sliding and 3D
diffusion for different DNA conformations on the search rate can be
found in Ref.~\cite{Firstpaper}.

\begin{figure}
\includegraphics[width=0.4\textwidth]{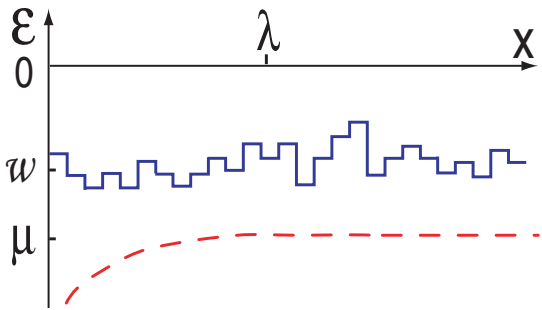}
\caption{Distribution of nonspecific adsorption energies $\epsilon$
and of chemical potential $\mu(x)$ along DNA molecule. The target
site is located at $x=0$, $\lambda$ is the antenna length. }
\label{fig:energy}
\end{figure}

Some authors calculate the typical time $\tau$ needed for the target
site to be found by a protein, when a small concentration $c$ of
proteins is randomly introduced into the system. Other
authors~\cite{Firstpaper} consider the specific site as a sink
consuming proteins with the diffusion limited rate $J$ proportional
to the concentration $c$ (which in turn should be supported on a
constant level by an influx of proteins into the system). Obviously
then, $\tau = 1/J$. Search rate enhancement due to the sliding along
DNA may be calculated as the ratio of the rate $J$ to the 3D
Smoluchowski rate $J_s = 4 \pi D_{3} c b$ of diffusion to the sphere
of radius $b$ modeling the target site on DNA. The central physical
idea is that one can define a piece of DNA adjacent to the target
for which 1D sliding diffusion dominates over parallel 3D diffusion
channel and which, therefore, serves as a receiving antenna for the
3D Smoluchowski-like diffusion of proteins. Then the key point of
the theory is to find the antenna length $\lambda$. In the language
of stationary flux $J$, this is done by matching incoming 3D flux
$J_3$ of proteins to the antenna with the 1D flux $J_1$ of proteins
sliding on the antenna toward the target.

All the cited above works assume that the nonspecific adsorption
energy $w$ of protein is sequence independent, i.e. the energy
profile experienced by the searching protein away from the target is
totally flat. This however disagrees with quasi-random character of
the natural sequences of DNA. It is known that the nonspecific
protein-DNA adsorption energy can be divided into two
parts~\cite{BH,Gerland}: (i) The sequence independent Coulomb energy
of attraction between the positively charged domain of the protein
surface and the negatively charged phosphate backbone, and (ii) the
sequence specific adsorption energy due to formation of hydrogen
bonds of the protein with the DNA bases. This is done by the
recognition $\alpha$-helix going deep into the major groove of
DNA~\cite{Gann}. Suppose the protein encounters $l$ base pairs
between positions $i$ and $i+l$. We call this position of the
protein \emph{site i} and characterize it by energy $\epsilon_i <
0$, where the energy of the free protein in water is chosen to be 0.
Because the sliding protein has a complex nonuniform structure and
interacts with a random DNA sequence, the total energy $\epsilon_i$
randomly fluctuates along DNA (Fig. \ref{fig:energy}). One can
assume that at nonspecific positions on DNA, the protein exploits
the same set of potential hydrogen bonds it forms with the
target~\cite{Barbi}. Since target recognition is often mediated by
hydrogen bonds to some of the four chemical groups on the major
groove side of the base pair~\cite{Nadassy}, and the recognition
$\alpha$-helix interacts with several base pairs, many hydrogen
bonds contribute to $\epsilon_i$. Therefore the distribution of
$\epsilon_i$ can be approximated by the Gaussian
distribution~\cite{Barbi,Mirny,Kardar} with a mean $w$ and standard
deviation $\sigma \ll |w|$:
\be g(\epsilon_i) =
\frac{1}{\sqrt{2\pi\sigma^2}}\exp\left[-\frac{(\epsilon_i-w)^2}{2\sigma^2}\right].
\label{eq:energy} \ee
In this paper we study a role of disorder on the rate enhancement
$J/J_s$ assuming that disorder is strong, i.e. $\sigma > kT$, where
$k$ is the Boltzmann constant and $T$ is the ambient temperature.

Similar to the the case of the flat energy
profile~\cite{Firstpaper}, we assume that transport outside the
antenna is \emph{mainly} due to the 3D diffusion, while inside the
antenna transport is \emph{dominated} by sliding, or 1D diffusion
along DNA and we equate the fluxes $J_1$ and $J_3$ to find
$\lambda$. The rate $J_3$ is given by the Smoluchowski formula for
the target size $\lambda$ and for the concentration of ``free'' (not
adsorbed) proteins $c_3$, it is $J_3 \sim D_3 c_3 \lambda$. The flux
on antenna $J_1$ strongly depends on $\sigma$ and also, generally
speaking, on DNA sequence in the finite antenna. We show that there
is a characteristic length of antenna $\lambda =
\lambda_c(\sigma,T)$ such that at $\lambda > \lambda_c$ flux $J_1$
self-averages and becomes sequence independent. Such a "macroscopic"
antenna determines $J/J_s$ for moderate disorder. In this case, the
ratio $J/J_s$ decreases exponentially fast with growth of disorder.
At stronger disorder we deal with a mesoscopic antenna with $\lambda
< \lambda_c$ and strictly speaking $J/J_s$ depends on random DNA
sequence. In this paper, we concentrate only on the most probable
value of $J/J_s$. In order to calculate it, we estimate the most
probable value of $J_1$. We show that in such a mesoscopic situation
disorder leads to a weaker reduction of $J/J_s$.

We assume that within some volume $v$ there is a straight, immobile
(double helical) DNA with the length $L$ smaller than $v^{1/3}$, but
much larger than any antenna length. For a dilute DNA solution,
$1/v$ stands for the concentration of DNA. We also assume that all
the microscopic length scales such as the length of a base pair, the
size of the target site, the diameter of the DNA etc. are of the
same order $b$. We are mainly interested in scaling dependence of
the rate enhancement $J/J_s$ on major system parameters, such as
$\sigma$, $w$, $L$ and $v$. This means that all the numerical
coefficients are dropped in our scaling estimates.

To estimate $J_1$, we assume at each site $i$ on DNA, the protein
has some probabilities of hopping to nearest neighboring sites $j$.
We write the probability for the hopping from an occupied site $i$
to an empty site $j$ as
\bea \gamma_{ij} \label{eq:hopping} &=&
\nu_0\exp\left(-\frac{\epsilon_{j}-\epsilon_i+|\epsilon_{j}-\epsilon_i|}{2kT}\right)
\nonumber \\ &=& \left\{\begin{array}{lcr} \nu_0\exp(-\frac{\epsilon_j-\epsilon_i}{kT}) & {\rm if} & \epsilon_j > \epsilon_i \\
\nu_0 & {\rm if} & \epsilon_j < \epsilon_i
\end{array} \right. ,\
\eea
where $\nu_0 \sim D_1/b^2$ is the effective attempt frequency. In
Eq. (\ref{eq:hopping}) we neglected the activation barriers
separating two states in comparison with $\epsilon_j - \epsilon_i$.
The number of proteins making such transition from site $i$ to $j$
per unit time can be estimated by
$\Gamma_{ij}=\gamma_{ij}f_i(1-f_j)$, where function $f_i$ is the
average occupation number of site $i$. At small enough $c$, all $f_i
\ll 1$ and thus $\Gamma_{ij} \simeq \gamma_{ij}f_i$. Function $f_i$
is given then by:
\be f_{i}=\exp[-(\epsilon_{i}-\mu_{i})/kT], \ee
where $\mu_{i}$ is the chemical potential. Using $\Gamma_{ij}$ and
$\Gamma_{ji}$, we can write the net flux from site $i$ to $j$ in the
form:
\be J_{ij}=\Gamma_{ij}-\Gamma_{ji} \simeq
\nu_0e^{-\frac{\epsilon_{ij}}{kT}}(e^{\frac{\mu_i}{kT}}-e^{\frac{\mu_j}{kT}}),\label{eq:flux}
\ee
where $\epsilon_{ij}=max\{\epsilon_i, \epsilon_j\}$.

We now argue that as long as the antenna is only a small part of the
DNA molecule, every protein adsorbs to DNA and desorbs many times
before it locates the target. Therefore, outside the antenna there
is statistical equilibrium between adsorbed and desorbed proteins,
and hence proteins have uniform chemical potential $\mu_i = \mu =
kT\ln(c_3b^3)$. Within the antenna, $\mu_i$ decreases when the site
approaches the target and reaches $-\infty$ at the target site (see
Fig. \ref{fig:energy}). If we label the border of the antenna as
site $1$ and the target as site $\lambda/b+1$, using Eq.
(\ref{eq:flux}), we can write
\be \sum_{i=1}^{\lambda/b}
J_{ij}e^{\frac{\epsilon_{ij}}{kT}}=
\nu_0(e^{\frac{\mu}{kT}}-e^{-\frac{\infty}{kT}})=\nu_0c_3b^3,
\ee
where $j=i+1$. Since the 1D current $J_1$ towards the target is
the same at any antenna site, i. e. $J_{ij} = J_1$, we can find it
as
\be J_1 = \label{eq:J}\frac{\nu_0c_3b^3}{\sum_{i=1}^{\lambda/b}
\exp(\epsilon_{ij}/kT)}\simeq
\frac{\nu_0c_3b^3\sqrt{2\pi\sigma^2}}{(\lambda/b)\int_{-\infty}^{0}
d\epsilon_{ij}R(\epsilon_{ij})}, \ee
where $R(\epsilon_{ij})$ is given by
\begin{eqnarray}
\label{eq:R}R(\epsilon_{ij})&=&\sqrt{2\pi\sigma^2}g(\epsilon_{ij})\exp(\epsilon_{ij}/kT)
\\
&=& \exp
\left\{\frac{\sigma^2}{2(kT)^2}+\frac{w}{kT}-\frac{[\epsilon_{ij}-(w+\sigma^2/kT)]^2}{2\sigma^2}\right\}\nonumber
.
\end{eqnarray}
One can interpret Eq. (\ref{eq:J}) as the Ohm's law, where the
numerator plays the role of the voltage applied to antenna and
denominator is the sum of resistances of all pairs $(i,j)$ which are
similar to Miller-Abrahams resistances for the hopping transport of
electrons~\cite{ES}.

The sharp maximum value of function $R(\epsilon_{ij})$ determining
the sum of Eq. (\ref{eq:J}) is reached when $\epsilon_{ij} =
\epsilon_{opt} = w+\sigma^2/kT$, and $R(\epsilon_{opt}) \sim
\exp[\sigma^2/2(kT)^2+w/kT]$. Thus
\be J_1 \sim
\frac{D_3c_3b^2}{\lambda}\exp\left[\frac{|w|}{kT}-\frac{\sigma^2}{2(kT)^2}\right],
\label{eq:JMCD}\ee
where we assumed for simplicity that $D_3= D_1 \sim b^2\nu_0$.

Before we move forward, we emphasize the crucial assumption already
made in above derivation. We assumed $\lambda$ is so long that
within the antenna the sliding protein encounters sites with energy
$\epsilon_{opt}$ more than once and therefore, the sum in Eq.
(\ref{eq:J}) can be replaced by the integral with limits from
$-\infty$ to $0$. We call such antenna macroscopic. For a short
antenna, the probability for such a site to appear inside is very
small. Thus the sum in Eq. (\ref{eq:J}) is determined by the largest
value of $R(\epsilon_{ij})$ typically available within the antenna.
We call such antenna mesoscopic.

\emph{Macroscopic antenna}---We study macroscopic antenna first.
Using $J_1$ and $J_3$, our main \emph{balance} equation for the rate
$J$ reads
\be J \sim D_3 c_3 \lambda \sim \frac{D_3 c_3 b^2}{\lambda}\exp\left[\frac{|w|}{kT}-\frac{\sigma^2}{2(kT)^2}\right] \ . \label{eq:balance} \ee
Thus the antenna length $\lambda$ is obtained as
\be \lambda \sim
b\exp\left[\frac{|w|}{2kT}-\frac{\sigma^2}{4(kT)^2}\right].
\label{eq:antenna1} \ee

\begin{figure}
\includegraphics[width=0.4\textwidth]{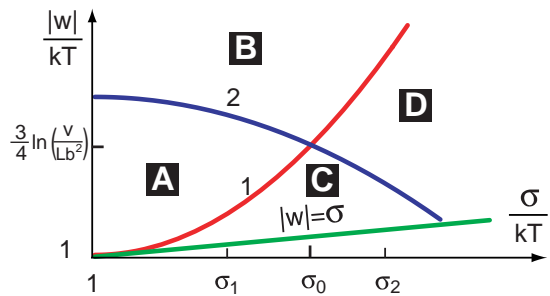}
\caption{(color online) The phase diagram of scaling regimes for
$|w|>\sigma>kT$. Each line marks a smooth crossover between scaling
regimes. The red line $|w| = 3\sigma^2/2kT$ marks the border 1
between macroscopic regimes (A, B) and mesoscopic regimes (C, D).
The blue line $|w|=kT\ln(v/Lb^2)-\sigma^2/2kT$ marks the border 2
between weak and strong adsorption regimes. They intersect at
$\sigma_0=kT[(1/2)\ln(v/Lb^2)]^{1/2}$, $|w|=kT(3/4)\ln(v/Lb^2)$.}
\label{fig:phase}
\end{figure}

Next we calculate the free protein concentration $c_3$. Suppose the
one-dimensional concentration of non-specifically adsorbed proteins
is $c_1$. Assuming the antenna is only a small part of the DNA and
remembering that adsorbed proteins are confined within distance of
order $b$ from the DNA, we can write down the equilibrium condition
as:
\be \frac{c_1}{c_3b^2} \sim \int f(\epsilon)
e^{-\epsilon/kT}d\epsilon \sim
\exp\left[\frac{|w|}{kT}+\frac{\sigma^2}{2(kT)^2}\right], \ee
which must be complemented by the particle counting condition $c_1L
+ c_3(v - Lb^2) = cv$. Since volume fraction of DNA is always small,
$Lb^2 \ll v$, standard algebra then yields
\bea c_3 & \simeq & \frac{cv}{yLb^2+v} \sim \left\{\begin{array}{lcr} c & {\rm if} & y < v/Lb^2 \\
cv/Lb^2y & {\rm if} & y > v/Lb^2
\end{array} \right. , \label{eq:equilibrium_concentrations} \eea
where $y$ is $\exp[|w|/kT+\sigma^2/2(kT)^2]$. Eqs.
(\ref{eq:equilibrium_concentrations}) lead to two different scaling
regimes, which are denoted as A and B in the diagram Fig.
\ref{fig:phase}. In regime A, the non-specific adsorption is
relatively weak, $c_3\sim c$, we arrive at
\be \frac{J}{J_s} \sim
\exp\left[\frac{|w|}{2kT}-\frac{\sigma^2}{4(kT)^2}\right].\ \ \ \ \
({\rm regime \ A}) \label{eq:regimeA}\ee
In the regime B, most proteins are adsorbed. Using the lower line of
Eqs. (\ref{eq:equilibrium_concentrations}), we obtain
\be \frac{J}{J_s} \sim
\frac{v}{Lb^2}\exp\left[-\frac{|w|}{2kT}-\frac{3\sigma^2}{4(kT)^2}\right].\
\ \ \ \ ({\rm regime \ B}) \label{eq:regimeB}\ee
In both regimes, $|w| > \sigma^2/kT$, thus $\sigma$ term of
$\ln(J/J_s)$ constitutes a correction. The size of antenna grows
with $|w|$, however unproductive non-specific adsorption of proteins
on distant pieces of DNA, which can slow down the transport to the
specific target grows with $|w|$ too. These two effects compete, as
a result the rate enhancement $J/J_s$ grows with $w$ in regime A and
declines in regime B. On the other hand, growing $\sigma$ reduces
the antenna size and promotes non-specific adsorption. Therefore,
$J/J_s$ decreases with $\sigma$ in both regimes.

The above theory deals with a macroscopic antenna. To be
macroscopic, the antenna has to contain at least one site with
energy around $\epsilon_{opt}$. The number of sites $n(\epsilon)$
with energy $\epsilon$ within the antenna is of the order of $\sim
(\lambda/b)\exp[-(\epsilon-w)^2/2\sigma^2]$. Thus a macroscopic
antenna requires $n(\epsilon_{opt})>1$, which gives $\lambda >
\lambda_c = b\exp[\sigma^2/2(kT)^2]$. Since we know $\lambda$ from
Eq. (\ref{eq:antenna1}), this condition can be written explicitly as
$|w| > 3\sigma^2/2kT$. Hence, $|w| = 3\sigma^2/2kT$ is the border
between the macroscopic regimes (A, B) and mesoscopic regimes (C, D)
in Fig. \ref{fig:phase}. We can check that when $|w|>3\sigma^2/2kT$,
the condition $\epsilon_{opt}< 0$ is satisfied for the case of
macroscopic antenna. Now we are ready to switch to the case of
mesoscopic antenna and explain regimes C and D.

\begin{figure}
\includegraphics[width=0.4\textwidth]{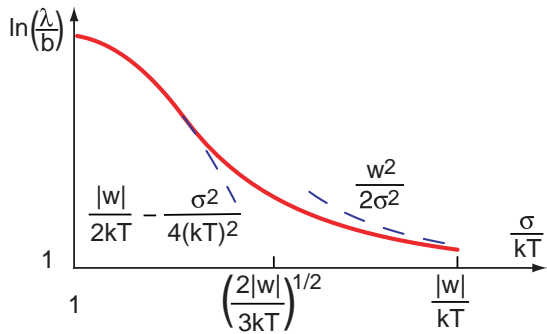}
\caption{Dependence of antenna length $\lambda$ on the disorder
strength $\sigma$. Dashed lines represent the asymptotic limits.}
\label{fig:lambda}
\end{figure}

\emph{Mesoscopic antenna}---In this case, the upper limit of the
integral in Eq. (\ref{eq:J}) should be replaced by
$\epsilon_{\lambda} \ll \epsilon_{opt}$ which is the largest energy
typically available within the antenna. It can be estimated from
$n(\epsilon_{\lambda}) \sim 1$, it is $\epsilon_{\lambda}\sim
w+\sqrt{2}\sigma\sqrt{\ln(\lambda/b)}$. Using $w$ and
$\epsilon_{\lambda}$, we can estimate the sum in Eq. (\ref{eq:J})
and get typical 1D current for the case of mesoscopic antenna:
\be \label{eq:JMSD}J_1(\lambda) \sim
D_3c_3b\exp\left[\frac{|w|}{kT}-\sqrt{2\ln(\lambda/b)}\frac{\sigma}{kT}\right].\ee
Eq. (\ref{eq:JMSD}) is apparently different from Eq. (\ref{eq:JMCD})
valid for the macroscopic antenna. This difference is partially
related to the rate enhancement of 1D diffusion at small time scale
noticed for the Gaussian disorder in computer
simulations~\cite{Barbi}. Equating $J_1(\lambda)$ to $J_3\sim
D_3c_3\lambda$, we obtain the antenna length
\be \lambda \sim
b\exp\left[\left(\sqrt{\frac{|w|}{kT}+\frac{\sigma^2}{2(kT)^2}}-\frac{\sigma}{\sqrt{2}kT}\right)^2\right].
\label{eq:antenna2} \ee
We can check, with this $\lambda$, that the condition
$\epsilon_{\lambda}<0$ still holds. When $|w|<\sigma^2/2kT$, the
antenna length $\lambda \sim b\exp(w^2/2\sigma^2)$. For a given
adsorption energy $w$, dependence $\lambda(\sigma)$ is plotted in
Fig. \ref{fig:lambda}. It shows that the decrease of the antenna
length with growing disorder strength slows down when antenna
becomes mesoscopic.

\begin{figure}
\includegraphics[width=0.4\textwidth]{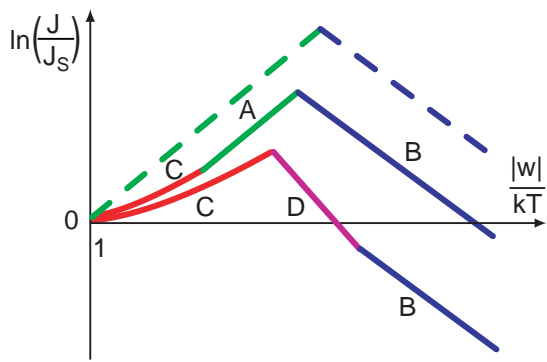}
\caption{Schematic plot of the dependencies of the rate enhancement
$J/J_s$ on $|w|$ at $\sigma=\sigma_1$ (upper solid curve) and
$\sigma=\sigma_2$ (lower solid curve). Letters A, B, C, D represent
the domains of Fig.\ref{fig:phase} they go through. Dashed line
shows the limit case of the flat energy profile with $\sigma=0$.}
\label{fig:rate}
\end{figure}
The crossover from a relatively weak adsorption to a strong one
described by Eqs. (\ref{eq:equilibrium_concentrations}) again leads
to the two scaling regimes for the case of mesoscopic antenna. They
are labeled C and D in the diagram Fig. \ref{fig:phase}. For
relatively weak adsorption, when $|w|<\sigma^2/kT$, we obtain regime
C, where
\be \frac{J}{J_s} \sim \exp\left(\frac{w^2}{2\sigma^2}\right),\ \ \
\ \ ({\rm regime \ C})\ee
while for strong adsorption we have regime D where
\be \frac{J}{J_s} \sim
\frac{v}{Lb^2}\exp\left[-\frac{|w|}{kT}-\frac{\sigma^2}{2(kT)^2}\right].\
\ \ \ \ ({\rm regime \ D}) \ee

In experiment, the adsorption energy $w$ can be controlled by the
salt concentration changing the Coulomb part of protein-DNA
interaction~\cite{BWH2}. The dependencies of $\ln(J/J_s)$ on $|w|$
at the two specified values of disorder strength $\sigma_1$ and
$\sigma_2$ marked in Fig. \ref{fig:phase} are schematically plotted
in Fig. \ref{fig:rate}. For comparison, we also plotted the case of
the flat energy profile ($\sigma=0$). In both cases with $\sigma
>0$, $\ln(J/J_s)$ first grows proportional to $w^2$ (regime C), because the antenna is mesoscopic and
thus 1D diffusion is faster, when compared to the normal diffusion
at macroscopic antenna. For a relatively small disorder
$\sigma=\sigma_1$, this rate enhancement continues to regime A but
with a rate proportional to $|w|$ because the antenna grows to be
macroscopic. For a larger disorder $\sigma=\sigma_2$, strong
nonspecific adsorption of proteins on distant pieces of DNA slows
down the search rate, when the antenna is still mesoscopic, and
$\ln(J/J_s)$ decreases in regime D faster than it does in regime B.
The antenna in regime B is macroscopic and $\ln(J/J_s)$ decreases
proportional to $|w|$ for both $\sigma=\sigma_1$ and
$\sigma=\sigma_2$.

The crossover from the weak disorder to the strong one happens at
$\sigma \sim \sigma_0=kT[(1/2)\ln(v/Lb^2)]^{1/2}$ (see Fig.
\ref{fig:phase}). If one plugs in the achievable experimental
conditions with $L/b \sim 150$ and $v\sim L^3$, estimate of
$\sigma_0$ is the order of $2kT$, which falls in the range of
estimates of $\sigma$ from $1kT$ to $6kT$ used in the
Refs.~\cite{Barbi,Mirny,Kardar}. Apparently $\sigma$ grows for
proteins with larger number of contacts with DNA and $\sigma_0$
decreases with DNA concentration. In order to identify the role of
strong disorder, we look forward to more experiments dealing with
relatively large concentrations of short straight DNA to guarantee
that disorder strength satisfies $\sigma>\sigma_0$.

We know only one observation ~\cite{BWH2} of the peak in the
coordinates of Fig. \ref{fig:rate} but for a long and definitely
coiled DNA for which our theory is not directly applicable. Indeed,
in this paper, we concentrated on the case of relatively short and,
therefore, straight DNA. In our recent paper~\cite{Firstpaper}, we
presented a general theory including Gaussian coiled and globular
DNA in the absence of disorder. In current paper, we did not touch
these cases because of our prejudice that simple questions should be
addressed first. We concentrated on the simplest regimes labeled A
and D in figure 4a of Ref.~\cite{Firstpaper} and still got rather
complicated diagram Fig. \ref{fig:phase} \footnote{We assume
$D_3=D_1$ in the absence of disorder. Thus with disorder, $d=D_1/D_3
< 1$ which corresponds to the case represented by the figure 4a of
Ref.~\cite{Firstpaper}}. That is why we did not try to present our
theory for more complicated regimes here.

We are grateful to A.Yu. Grosberg, S.D. Baranovskii and J. Zhang for
useful discussions.


\begin{thebibliography}{99}

\bibitem{Gann} M. Ptashne and A. Gann, \emph{Genes and signals} (Cold Spring Harbor Laboratory Press, Cold Spring Harbor, NY,
2001).
\bibitem{Riggs} A.D. Riggs, S. Bourgeois, and M. Cohn. J. Mol. Bol. {\bf 53}, 401 (1970).

\bibitem{Richter} P.H. Richter and M. Eigen. Biophys. Chem. {\bf 2}, 255 (1974).

\bibitem{BWH} O.G. Berg, R.B. Winter and P.H. von Hippel. Biochemistry {\bf 20}, 6929 (1981).

\bibitem{Bruinsma} R.F. Bruinsma. Physica A {\bf 313}, 211 (2002).

\bibitem{Marko} S.E. Halford and J.F. Marko. Nucleic Acids Res. {\bf 32}, 3040 (2004).

\bibitem{Halford} S.E. Halford and M.D. Szczelkun. Eur. Biophys. J. {\bf 31}, 257 (2002).

\bibitem{Klenin} K.V. Klenin, H. Merlitz, J. Langowski, C.X.
Wu, Phys. Rev. Lett. {\bf 96}, 018104 (2006).

\bibitem{Firstpaper} Tao Hu, A. Yu. Grosberg and B. I. Shklovskii.
Biophys. J. {\bf 90}, 2731 (2006).

\bibitem{BH} O.G. Berg and P.H. von Hippel. J. Mol. Biol. {\bf 193}, 723 (1987).

\bibitem{Gerland} U. Gerland, J.D. Moroz and T. Hwa. Proc.
Natl. Acad. Sci. USA. {\bf 99}, 12015 (2002).

\bibitem{Barbi} M. Barbi, C. Place, V. Popkov and M. Salerno. Phys. Rev. E {\bf 70}, 041901 (2004);
J. Biol. Phys. {\bf 30}, 203 (2004).

\bibitem{Nadassy} K. Nadassy, S.J. Wodak and J. Janin. Biochemistry
{\bf 38}, 1999 (1999).

\bibitem{Mirny} M. Slutsky and L.A. Mirny. Biophys. J. {\bf 87}, 4021 (2004).

\bibitem{Kardar} M. Slutsky, M. Kardar and L.A. Mirny. Phys. Rev. E {\bf 69}, 061903
(2004).

\bibitem{ES} B.I. Shklovskii and A.L. Efros, \emph{Electronic Properties of Doped
Semiconductors} (Springer-Verlag, Berlin, 1984).

\bibitem{BWH2} R.B. Winter, O.G. Berg, and P.H. von Hippel. Biochemistry {\bf 20}, 6961 (1981).



\end{thebibliography}
\end{document}